\documentclass[twocolumn,superscriptaddress,secnumarabic,amssymb,nobibnotes,aps,prb]{revtex4-2}

\usepackage{extarrows}
\usepackage{amsmath}    
\usepackage{graphicx}   
\usepackage{verbatim}   
\usepackage{color}      
\usepackage{subfigure}  
\usepackage{hyperref}   
\usepackage{verbatim}   
\usepackage{longtable}
\usepackage{epsfig}
\usepackage{comment}
\usepackage{bm}
\usepackage{dcolumn} 
\usepackage{textcomp}   
\usepackage{kotex}      
\newcommand{\css}{CrSbS$_3$}
\newcommand{\csse}{CrSbSe$_3$}

\begin{document}
\title{Bond-Length–Driven Magnetic Transition\\
in Quasi-One-Dimensional CrSb$X_3$ ($X$=S, Se)
} 
\author{Kang Lee}
\affiliation{Department of Applied Physics, Graduate School, Korea University, Sejong 30019, Korea}
\author{Hong-Suk Choi}
\affiliation{Department of Applied Physics, Graduate School, Korea University, Sejong 30019, Korea}
\author{Kwan-Woo Lee}
\email{mckwan@korea.ac.kr}
\affiliation{Department of Applied Physics, Graduate School, Korea University, Sejong 30019, Korea}
\affiliation{Division of Semiconductor Physics, Korea University, Sejong 30019, Korea}

\begin{abstract}
Using {\it ab initio} calculations, we investigate the magnetic ground states of quasi-one-dimensional insulating CrSb$X_3$ ($X$ = S, Se) with infinite double-rutile chains. Within conventional band theory, without explicit Coulomb correlations ($U$), we obtain band gaps in close agreement with experiment. Remarkably, we find that the magnetic order is highly sensitive to the Cr–Cr bond length $d_{\rm Cr-Cr}$: increasing the bond length induces a transition from antiferromagnetic to ferromagnetic order at a critical distance $d^c_{\rm Cr-Cr} \approx 3.53 (\pm 0.05)$ \AA.  
Accordingly, CrSbS$_3$ lies near the transition boundary, whereas CrSbSe$_3$ is robustly ferromagnetic, in good agreement with experiment.
Analysis of the exchange interactions reveals that the first-order phase transition is dominated by a sign reversal of the intrachain nearest-neighbor superexchange $J_1$ mediated by chalcogen ions, while the intrachain direct exchange $J_2$ remains ferromagnetic and changes only gradually. This behavior reflects an emergent Bethe–Slater-like behavior driven by competing exchange pathways in a quasi-1D transition-metal system, where the competition between $J_1$ and $J_2$ dictates the magnetic ground state.
Besides, the electronic structures of the ground states of each compound are investigated.
\end{abstract}

\maketitle
\clearpage

\section{Introduction}
Low-dimensional systems host a variety of intriguing phenomena, including Peierls instabilities, spin- or charge-density waves, exotic magnetism, and superconductivity.
In particular, according to the Mermin–Wagner theorem \cite{mermin}, a purely one- or two-dimensional (1D/2D) system cannot sustain spontaneous magnetization at finite temperature. However, quasi-1D and -2D systems often manifest magnetic states due to weak inter-chain interactions or coupling along other crystallographic directions, often called the van der Waals magnet \cite{jg.park,s.yang,klein}.
Interestingly, the quasi-1D ternary chromium trichalcogenides CrSb$X_3$ ($X$=S, Se),
which consist of the double-rutile chain structure of Cr$X_6$ octahedra extending along the $\hat{b}$-axis \cite{odink,volkov}, 
show distinct, yet sometimes controversial, magnetic states, {\it i.e.}, insulating ferromagnetic (FM) versus antiferromagnetic (AFM) orders \cite{odink,volkov,cava2018,c.li2025}.

\csse~is a rare insulating FM of an optical energy gap $E_g\sim$0.7 eV and the Curie temperature $T_C$=71 K, showing the saturated moment 3$\mu_B$/Cr \cite{odink,cava2018,y.sun2020,y.qu2021}, 
consistent with a Cr$^{3+}$ ($3d^3$) ionic configuration of spin $S=\frac{3}{2}$.
This compound shows a large anisotropic magnetocaloric effect  \cite{cava2018,y.sun2020,y.liu2020}, with the easy axis $\hat{a}$ perpendicular to the double-rutile chain, as theoretically confirmed \cite{mathew,y.xun2021,g.wang2021}.
The magnetocrystalline anisotropy constant is $K_u\sim$145 kJ/m$^3$ at 10 K and decreases by $\sim$73\% at $T_C$ for the hard axis $\hat{b}$ \cite{y.liu2020}.
In contrast, the $T$ dependence of $K_u$ along the other crystallographic directions is comparatively weak.
Based on {\it ab initio} calculations,
a very low thermal conductivity of 0.56 W/m$\cdot$K was suggested, 
leading to a high thermoelectric figure of merit 0.46 at 900 K upon Pb doping \cite{d.yang2017}.
A previous theoretical investigation suggested that, based on the calculated moderate cleavage energy, 
a \csse~ladder structure could be exfoliated, with a predicted $T_C$ more than twice that of the bulk \cite{y.xun2021}.
Besides, applying pressure $P$, the insulating FM \csse~undergoes a gradual transition into a superconducting state, with the onset occurring at $P_c\approx33$ GPa. 
The superconductivity reaches the maximum critical temperature of $T_c$=7.7 K at $P_c\approx58$ GPa \cite{c.li2024}.

In contrast, although \css~was initially reported to be an insulating FM \cite{odink,volkov}, a recent experiment shows an AFM order \cite{c.li2025}, implying that its magnetic state is sensitive.
The resistivity measurement indicates an insulating
state of $E_g\approx0.9$ eV, confirmed by the heat capacity measurement \cite{comment1}.
The susceptibility data show two peaks at $T_{CT}$=94 K and the N\'eel temperature $T_N$=43 K, accompanying a kink and a hump in the specific heat measurements, respectively. 
The measurements of neutron powder diffraction indicate
that \css~shows a C-type AFM, featuring a zigzag spin alignment along each double rutile chain, the easy axis $\hat{a}$ as in \csse, and no spin canting.
The long-range spin order was confirmed by asymmetric features in the electron spin resonance (ESR) spectra below $T_N$.
The observed saturated local moment of the Cr ion is 2.37 $\mu_B$, much smaller than the nominal value of $S=\frac{3}{2}$,
implying substantial $p-d$ hybridization. 
Besides, seemingly at around 50 GPa, an insulator-to-metal transition appeared, 
but no superconductivity was observed.

Here, we investigate the electronic structures and the origin of the distinct magnetic states in these quasi-1D insulating \csse~and \css, using {\it ab initio} calculations.
Our results reveal a first-order magnetic phase transition from FM to AFM order upon shortening the Cr–Cr bond length $d_{\rm Cr-Cr}$, reminiscent to the Bethe–Slater curve \cite{slater,goldman} that accounts for the emergence of FM or AFM in elemental transition metals in terms of interatomic distance.
Recently, this curve was analyzed by scrutinizing variations of the orbital-decomposed nearest-neighbor (NN) exchange parameters in elemental $3d$ metals with the bcc or fcc structure \cite{kvash2016,kvash2017}.
By evaluating the magnetic exchange parameters in CrSb$X_3$, we show that the magnetic ground state is governed by a competition between the NN $J_1$ and next-nearest-neighbor (NNN) $J_2$ intra-chain interactions. 
Notably, $J_1$ is highly sensitive to $d_{\rm Cr-Cr}$, 
evolving from AFM to FM coupling as $d_{\rm Cr-Cr}$ increases, thereby driving the magnetic phase transition.
Besides, we will briefly analyze the electronic structures of both the underlying nonmagnetic state and the magnetic ground states of these systems that show strong 1D-characters.
Before summarizing, in the Discussion
we will shortly discuss the pressure-induced phase transition in \csse~and the unusual experimental observations in the intermediate $T$ region near $T_{CT}$ in \css.

\begin{figure}[t]
 \centering
\includegraphics[width=\columnwidth]{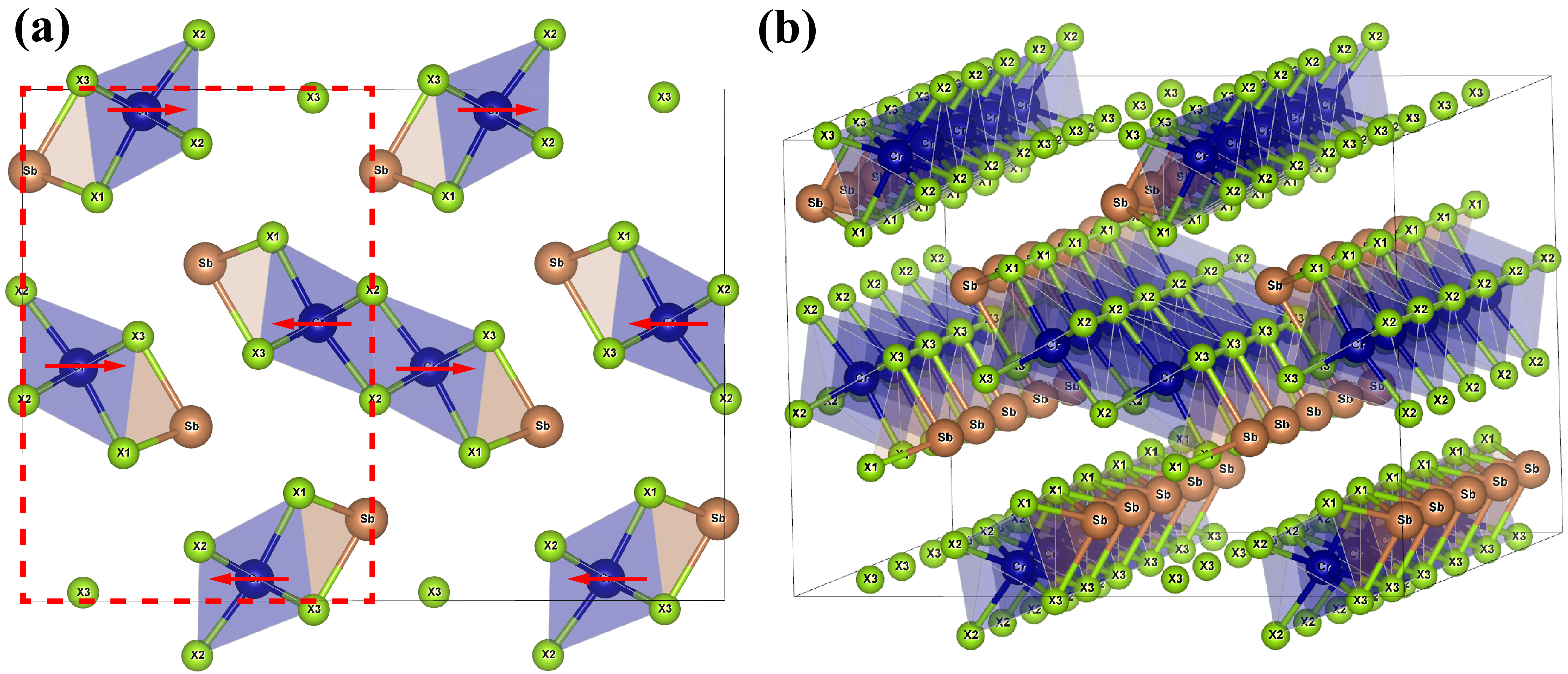}
\caption{Crystal structure of the quasi-1D CrSb$X_3$ ($X$=chalcogens) systems. (a) Top view in the $ac$ plane.
 The red-dashed line outlines the unit cell.
 The arrows indicate the spin directions for the AFM order with the easy axis $\hat{a}$ in \css.
 (b) Cr$X_6$ double rutile chains, sharing four common edges, along the $\hat{b}$-axis.
 The blue, brown, and green spheres indicate Cr, Sb, and $X$ (S or Se) ions, respectively.
 	}
	\label{str}
\end{figure}

\section{Crystal Structure and Calculation Approaches}
\label{method}
Figure \ref{str} shows the crystal structure of the ternary
chromium trichalcogenides CrSb$X_3$ (space group: $Pnma$, No. 62), 
containing 4 formula units (f.u.) in the unit cell. In these systems,
the edge-shared octahedra Cr$X_6$ form infinite
double-rutile chains that are linked by Sb$X_3$ pyramids
and extend along the $\hat{b}$-axis. The double-rutile chains are
weakly connected by the van der Waals interaction in the
$ac$ plane. The chains are oriented obliquely in this plane.

\begin{table}[tb]
\begin{center}
\caption{Information of crystal structures of CrSb$X_3$ ($X$=S, Se).
For the relaxation, obtained from unpolarized spin calculations, we used the experimental lattice parameters measured at room $T$ \cite{cava2018,c.li2025}: $a$= 8.6686 (9.1388), $b$=3.6194 (3.7836), $c$=12.8773 (13.3155) for \css~(\csse), in units of \AA.
(The differences between spin-polarized and unpolarized cases are small, just a few hundredths of an \AA.)  
In the space group $Pnma$, all atoms sit at the $4c$ sites.
For the experimental internal parameters, see text.
}
\label{table1}
\begin{tabular}{ccccccccc}\hline\hline
      \multicolumn{2}{c}{}&\multicolumn{3}{c}{Exp.}& ~ & \multicolumn{3}{c}{Relx.} \\\cline{3-5}\cline{7-9}
  compounds & atom & $x$ & $y$  & $z$ &~&  $x$ & $y$  & $z$ \\  \hline
  \css~   & Cr  &  0.1564 & 3/4 & 0.5374 &~& 0.1585 & 3/4 & 0.5445 \\
   ~   & Sb  & 0.4823 & 1/4 & 0.6603 &~& 0.4833 & 1/4 & 0.6620 \\
   ~ & S   &0.2883 & 3/4 & 0.7053 &~& 0.2900 & 3/4 & 0.7087 \\
   ~& S   &0.9948 & 3/4 & 0.3948 &~& 0.9985 & 3/4 & 0.3960 \\
   ~&S   &0.3322 & 1/4 & 0.4852&~& 0.3309 & 1/4 & 0.4892 \\\hline
  \csse~ & Cr  &  0.1555 & 3/4 & 0.5447 &~& 0.1591 & 3/4 & 0.5430 \\
   ~   & Sb  & 0.4706 & 1/4 & 0.6579 &~& 0.4797 & 1/4 & 0.6593 \\
   ~ & Se  &0.2848 & 3/4 & 0.7131 &~& 0.2870 & 3/4 & 0.7106 \\
   ~& Se   &0.0019 & 3/4 & 0.3911 &~& 0.0034 & 3/4 & 0.3963 \\
   ~&Se   &0.3280 & 1/4 & 0.4845&~& 0.3285 & 1/4 & 0.4845 \\
\hline\hline
\end{tabular}
\end{center}
\end{table}

Since van der Waals interactions are crucial for determining stable structures in such quasi-1D systems, we optimized the crystal structures of these compounds using the density functional theory plus dispersion (DFT-D3) method \cite{dft-d}, as implemented in the Vienna ab initio simulation package ({\sc vasp}) \cite{vasp1,vasp2}.
All atomic positions sitting at $4c$ sites $(x,\frac{1}{4},z)$ were fully optimized until the forces were less than 0.01 eV/\AA, 
with the lattice parameters experimentally observed  at room $T$ \cite{cava2018,c.li2025}. 
The results are given in Table \ref{table1}, 
with the experimental information for comparison.
These yield $d_{\rm Cr-Cr}$ (in units of \AA) of 3.39 (3.48) in \css~ and 3.60 (3.66) in \csse~for the experimental (optimized) structures.
(The calculated cohesive energies, $E_{\rm atom}-E_{\rm bulk}$, in the experimental structures are 4.32 eV/atom for \css~ and 3.83 eV/atom for \csse, confirming that both compounds are energetically bound.)
The experimental and optimized values are very close to each other, differing only in the hundredth digit of an \AA.
As will be addressed below, however,
$d_{\rm Cr-Cr}$ is a crucial factor for determining the magnetic ground state in these compounds. 
It should be noted that no experimentally measured internal parameters are available for \css. Therefore, to represent an experimental structure in \css, we reduced $d_{\rm Cr-Cr}$ by approximately 0.1 \AA~from our optimized value, based on the observation that \csse-for which experimental data are available-shows a similar trend.
With the $d_{\rm Cr-Cr}$ fixed at this adjusted value, 
the remaining internal parameters were optimized.
(Hereafter, we will refer to this as the experimental structure \css.)
This procedure was also utilized when $d_{\rm Cr-Cr}$ was varied (see below).
Note that our relaxations of the ionic internal parameters show that the local structures
remain nearly unchanged, 
even as the lattice parameters vary within the range of experimentally observed $T$-dependent variations \cite{c.li2025}.

With the optimized and experimental internal parameters,
our {\it ab initio} calculations, based on the revised Perdew-Burke-Ernzerhof generalized gradient approximation PBEsol (simply called GGA, here) \cite{pbe-sol}, were carried out with the all-electron full-potential {\sc wien2k} code \cite{wien2k}.
In contrast to previous calculations of \csse~\cite{g.wang2021} with the projector-augmented-wave code {\sc vasp}, which largely underestimated the energy gap $E_g$, our GGA calculations with the {\sc wien2k} code yield band gaps for both systems that are close to the experimental values.
(For details, see below.)
This gap issue is further corroborated by independent calculations using another all-electron full-potential code {\sc fplo} (not shown here) \cite{fplo}. 
The basis size in {\sc wien2k} was determined by $R_{mt}K_{max}$ = 7 and the augmented plane wave (APW) radii of Cr 2.33, Sb 2.5, S 2.01, and Se 2.33 (in units of a.u.).
The Brillouin zone (BZ) was sampled with a $k$-mesh of $8\times20\times5$ that is sufficient for such insulating systems.

To calculate the magnetic exchange parameters, we constructed
the maximally localized Wannier functions \cite{marzari97} 
with the {\sc wannier90} \cite{wan90} and {\sc wien2wannier} \cite{wien2wan} codes, projecting onto symmetry adapted Wannier functions.
In the frozen energy windows of $-6$ eV to 4 eV for starting orbitals of Cr $3d$, Sb and $X$ $p$ orbitals, 
the excellent agreement of the Wannier bands with the AFM bands are shown (see supplementary materials (SM)).
The obtained Wannier functions are symmetrized by using the {\sc wannsymm} code \cite{wansym}.
With the symmetrized Wannier functions, the magnetic exchange parameters were calculated using the {\sc tb2j} package \cite{tb2j}
in which the local rigid spin rotation is treated as a perturbation 
within the single-particle Green's function method \cite{liecht}.

\section{Results}

\begin{figure}[ht] 
 \centering
\includegraphics[width=0.50\textwidth]{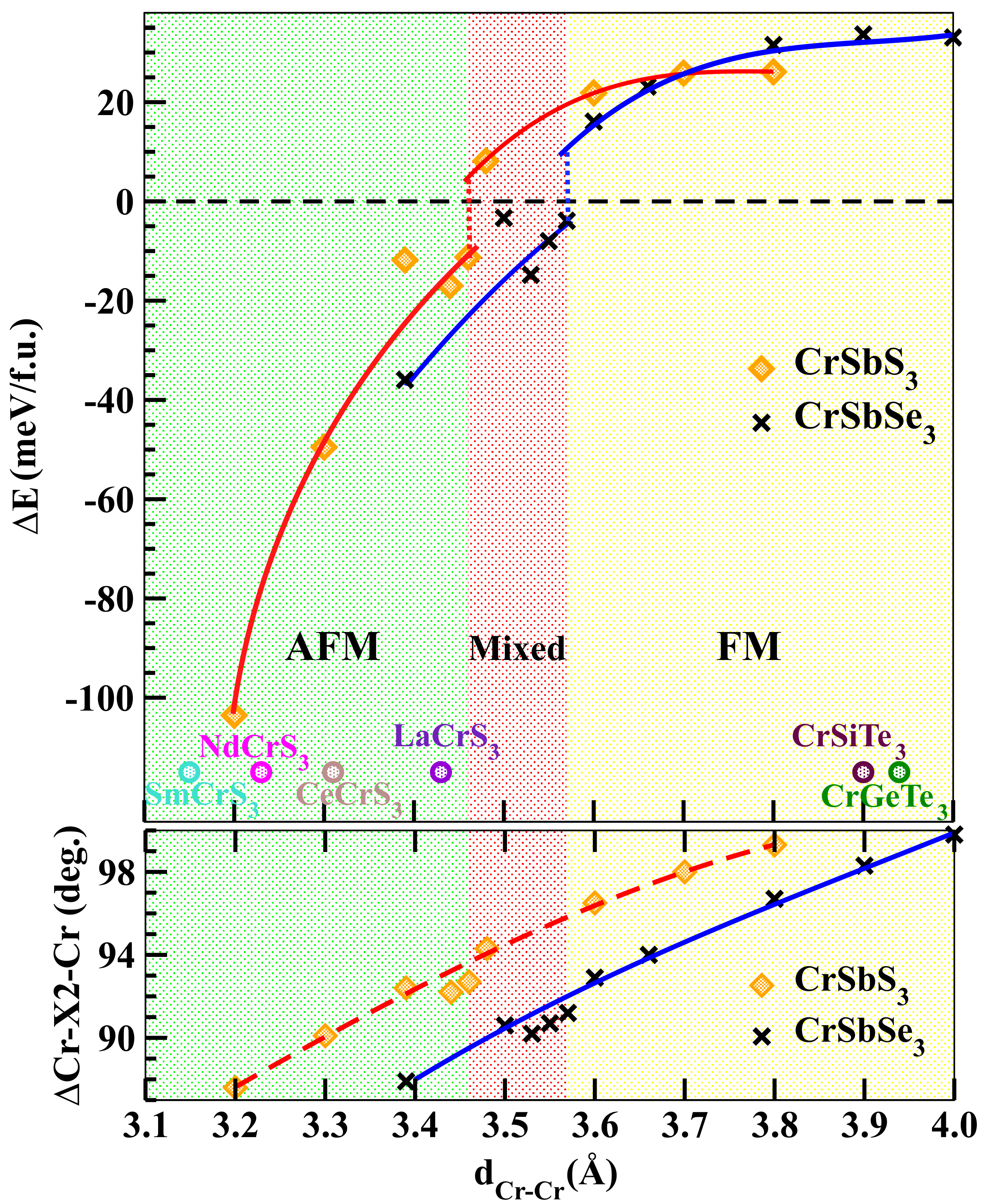}
\caption{Top: variations of energy difference $\Delta E_{\rm AFM-FM}$ between AFM and FM for both compounds, as changing the Cr-Cr bond length $d_{\rm Cr-Cr}$.
The variations, indicated by the red (blue) lines for \css~(\csse),
show approximately cubic variations, 
but evident discontinuities in the boundaries preceding the AFM–FM transition. 
In the mixed region of $3.48\lesssim d_{\rm Cr-Cr}\lesssim3.58$ (in units of \AA), FM is energetically favored over AFM in \css,
while AFM is favored in \csse.
The experimental bond lengths for four AFM compounds ${\cal R}$CrS$_3$ (${\cal R}$= Sm, Nd, Ce, and La) and two FM compounds CrGdTe$_3$ and CrSiTe$_3$ are adapted from Ref. \cite{kikk} and Ref. \cite{siber}, respectively.
Bottom: changes in Cr-$X2$-Cr bond angle, as varying $d_{\rm Cr-Cr}$,
showing roughly linear variations.
Here, $X_2$ denotes a chalcogen ion that mediates the intra-chain Cr–Cr superexchange interaction.
}
\label{energy}
\end{figure}

\subsection{Energetics}
\label{energetics}
To find the magnetic ground state for these compounds, we first investigated changes in energy, as varying $d_{\rm Cr-Cr}$ in the range of 3.2 \AA~ to 4.0 \AA, encompassing all possible values in these systems.
Throughout the entire range, in both of the compounds the nonmagnetic state is energetically unfavorable, lying approximately 1 eV/f.u. above the AFM and FM states,
indicating a strong tendency toward spin-orders.

Over the $d_{\rm Cr-Cr}$ range investigated here, 
for both systems the magnetic moments remain nearly unchanged,
reflecting the insulating character: the total moment of 3 $\mu_B$ in the FM and the Cr local moment of 2.7-2.8 $\mu_B$ in the AFM,
corresponding to a $S=\frac{3}{2}$ state.
The changes in energy of each magnetic state can be well fitted by a function including terms up to cubic order, reflecting the van der Waals interaction for the quasi-1D systems, 
but some deviations near the equilibrium positions.
As usual in the systems,
these changes show an asymmetric behavior about each equilibrium position. 
The compression case is stiffer than the stretching case, being about 6 times stiffer in \css~and twice stiffer in \csse~(for details, see SM). 

From these energy changes, we calculated the variations in the energy difference $\Delta E$ between the AFM and FM states for both compounds,
as varying $d_{\rm Cr-Cr}$.
As shown in the top panel of Fig. \ref{energy},
these variations show an approximately cubic decrease with decreasing $d_{\rm Cr-Cr}$, analogous to the energy changes of each magnetic state,
but pronounced discontinuities at the boundaries of the transition between AFM and FM.
This indicates a first-order phase transition of FM-AFM driven by variations in $d_{\rm Cr-Cr}$.
FM is favored over AFM for $d_{\rm Cr-Cr}\gtrsim3.58$ \AA~in \csse~and
for $d_{\rm Cr-Cr}\gtrsim3.48$ \AA~in \css.
This transition is analogous to the Bethe-Slater curve \cite{slater,goldman},
but with the critical value $d^c_{\rm Cr-Cr}\approx 3.53 (\pm 0.05)$ \AA. 
As given in the top panel of Fig. \ref{energy},
the magnetic ground states of the well-known quasi-1D Cr-based compounds, four AFM compounds ${\cal R}$CrS$_3$ (${\cal R}$= Sm, Nd, Ce, and La) and two FM compounds CrGdTe$_3$ and CrSiTe$_3$, are well distinguished by the $d^c_{\rm Cr-Cr}$ value,
indicating that this value is broadly applicable to the quasi-1D Cr-based compounds.
In 2D transition metal dichalcogenide bilayers of $MX_2$ ($M$=V, Cr, Mn; $X$=S, Se, Te), a similar trend was found when tuning the interlayer $X-X$ distance \cite{c.wang2020}.
The origin of the magnetic phase transition in \css~and \csse~will be discussed below.
Note that the magnetic ground states of \csse~and \css~based on the experimental structures are consistent with the previous theoretical results \cite{mathew}, indicating the robustness of our findings.

The bottom panel of Fig. \ref{energy} displays the variation of the bond angle Cr-$X2$-Cr with decreasing $d_{\rm Cr-Cr}$, showing an approximately linear reduction from 100$^\circ$ to 88$^\circ$,
albeit with some deviations near the mixed region.
So, variation in $d_{\rm Cr-Cr}$ is analogous to change in the bond angle.
Below $\sim$91$^\circ$ (93$^\circ$) for \csse~(\css), these compounds are AFM. This indicates that these systems do not closely follow the empirical Goodenough–Kanamori–Anderson rule predicting ferromagnetic coupling for 90$^\circ$ superexchange pathways in the half-filled systems such as these compounds.

Moreover, both compounds are insulating in these magnetic states already with the GGA, in this $d_{\rm Cr-Cr}$ range.
Using the experimental value of $d_{\rm Cr-Cr}$ in \csse, 
earlier calculations within the local spin density approximation (LSDA) or the GGA using {\sc vasp} also found an insulating ground state, but with a $E_g$ significantly smaller than the experimental value \cite{y.xun2021,g.wang2021}. 
To reproduce the observed $E_g$, those studies employed either the LSDA+U approach \cite{g.wang2021} or the Heyd–Scuseria–Ernzerhof (HSE) hybrid functional \cite{y.xun2021}, and consequently interpreted the system as a Mott insulator \cite{g.wang2021}.
In contrast, our calculations for both compounds
using {\sc wien2k}, and independently confirmed with {\sc fplo}, 
yield $E_g$'s already at the GGA level that are in close agreement with the experiment measurements \cite{cava2018,c.li2025}.
Both {\sc wien2k} and {\sc fplo} codes are all-electron full-potential methods, allowing an accurate description of the anisotropic potential and the localized $3d$ orbitals without relying on shape or pseudopotential approximations.
This indicates that the insulating behavior in these compounds can be understood within a band-insulating feature, without invoking strong correlation effects beyond GGA.

\begin{figure}[tb] 
\centering
\includegraphics[width=0.50\textwidth]{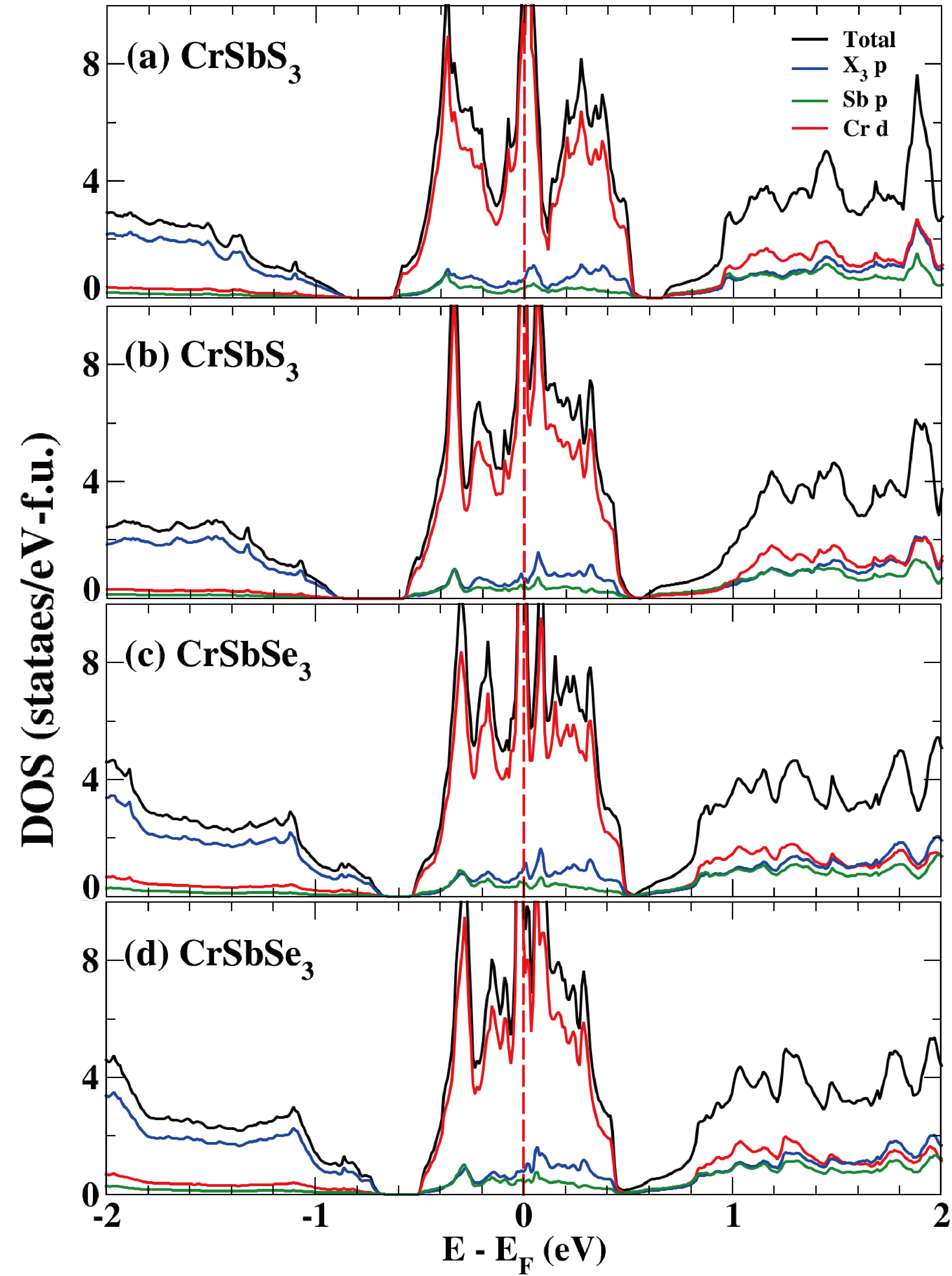}
\caption{Enlarged NM total and atom-projected densities of states (DOSs)
for $d_{\rm Cr-Cr}$=3.39, 3.48, 3.60, 3.66 \AA~ (from top to bottom) in \css~(a),(b), and \csse~(c),(d).
(a) and (c) represent the experimental $d_{\rm Cr-Cr}$,
while (b) and (d) correspond to our optimized $d_{\rm Cr-Cr}$.
These DOSs show a sharp peak at the Fermi energy $E_F$, characteristic in (quasi-)1D systems. 
}
\label{dos_nm}
\end{figure}

\subsection{Electronic structures}
\label{el_str}
In this subsection, we briefly discuss the underlying NM electronic structures of both compounds, with particular emphasis on chalcogen ion-dependent variations.
We then analyze the electronic structures of the FM \csse~and the AFM \css, which correspond to the 
ground states of their respective compounds.

{\it NM states of both compounds.}
Figure \ref{dos_nm} shows the total and atom-projected DOSs for the NM state of \css~and \csse~at the experimental and optimized $d_{\rm Cr-Cr}$.
The DOSs near the Fermi energy $E_F$, which have the character of the isolated $t_{2g}$ manifold, show roughly three peaks and an $E^{-\frac{1}{2}}$ dependence, consistent with quasi-1D characteristics \cite{sjkim}.
The half-filled $t_{2g}$ manifold, indicating a Cr$^{3+}$ ($3d^3$) configuration,
is separated from the filled $X$ $p$ orbital with the $p-d$ hybridization gap of at most $\sim$0.25 eV and from the unfilled $e_g$ manifold that extends in the range of approximately 0.5 eV to 4 eV. 
The crystal field splitting between the $t_{2g}$ and $e_g$ manifolds, measured from their centers, is about 1.8 eV.
As expected, the $t_{2g}$ manifold becomes more localized, with its bandwidth gradually decreasing from 1.1 eV to 0.9 eV as $d_{\rm Cr-Cr}$ increases from 3.39 \AA~to 3.66 \AA.
Additionally, the larger $p-d$ hybridization gap and increased $X$ $p$-orbital character in the $t_{2g}$ region indicate stronger Cr-$X$ covalent bonding in \css~compared to \csse, reflecting the shorter Cr-$X$ distance by $\sim$0.1 \AA~in average in \css.

\begin{figure}[tb] 
\centering
\includegraphics[width=\columnwidth]{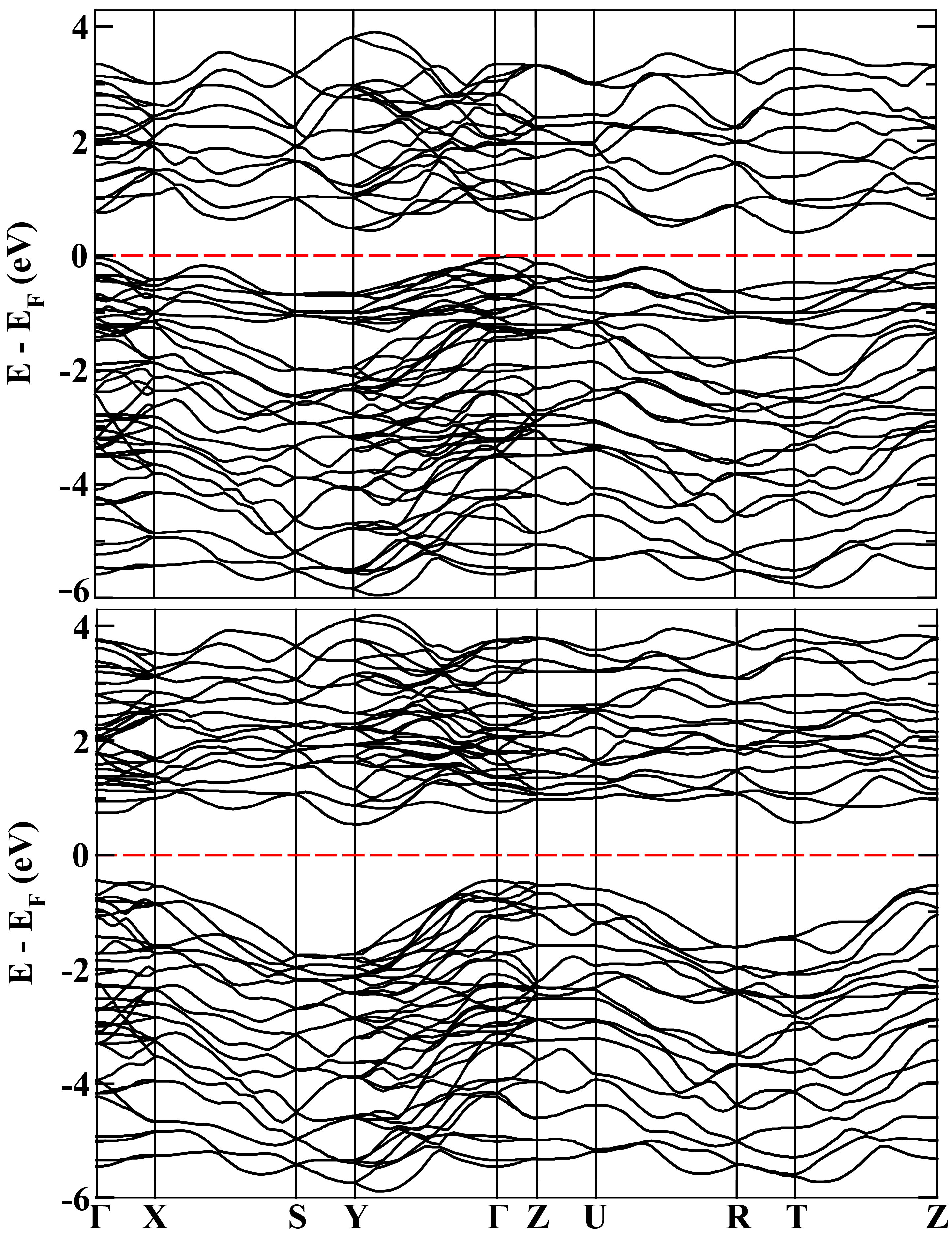}
\caption{FM band structure of \csse, in the range of $-6$ eV to 4 eV containing Cr $d$ and $p$ bands of the other ions, at the experimental crystal structure with $d_{\rm Cr-Cr}=3.60$ \AA.
(a) and (b) correspond to the majority and minority channels, respectively, indicating an insulating state.
Note that bands stick at the top of BZ along the $Z-U-R-T-Z$ line,
as commonly appeared in the non-symmorphic crystals.
}
\label{fm_band}
\end{figure}

\begin{figure}[tb] 
\centering
\includegraphics[width=\columnwidth]{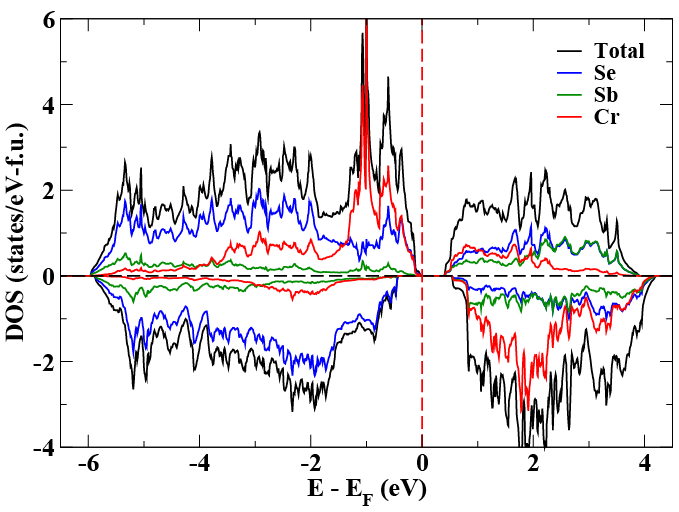}
\caption{FM total and atom-projected DOSs of \csse, at the experimental structure with $d_{\rm Cr-Cr}=3.60$ \AA, indicating an insulating state of $E_g\sim0.5$ eV and $t_{2g}^{3\uparrow}$.
}
\label{fm_dos}
\end{figure}

{\it FM state of \rm \csse.}
For the compound, the FM state is favored over the AFM state for $d_{\rm Cr-Cr}\gtrsim3.58$ \AA, which lies within the range of both experimentally observed and theoretically optimized $d_{\rm Cr-Cr}$ values,
consistent with the experimental observations \cite{odink,cava2018,y.sun2020,y.qu2021}.
The energy difference is about 20 (15) meV/f.u. for the optimized (experimental) $d_{\rm Cr-Cr}$,
corresponding to an estimated superexchange $J'\approx4.4 (3.3)$ meV$\approx50$ K from a simple Ising picture for a 1D system of $S=\frac{3}{2}$.
This value is comparable to the experimental $T_{C}$.

The FM band structure of \csse~is given in Fig. \ref{fm_band}, at the experimental crystal structure with $d_{\rm Cr-Cr}=3.60$ \AA~\cite{cava2018}.
The corresponding total and atom-projected DOSs are shown in Fig. \ref{fm_dos}.
In the majority channel,
the filled Cr $d$ orbitals exist mostly in the region above $-1.4$ eV and the unfilled $e_{g}$ manifold lies in the bottom of the conduction band,
while the Se $p$ orbitals spread in the range of $-6$ eV to $-1.2$ eV with some mixing with the filled Cr $d$ orbitals in the region above $-1.2$ eV.
The $d-d$ $E_g$ of about 0.5 eV is formed,
close to the experimental value of 0.6-0.7 eV \cite{cava2018}.
The top of the valence band shows pronounced quasi-1D character, with a relatively strong dispersion of $\sim$1.2 eV along the $\Gamma-Y$ line,
yielding an estimated NN hopping parameter along the $\hat{b}$-axis of $t_b\approx0.3$ eV. 
In the minority channel, the filled Se $p$ orbitals exist in the range of $-6$ eV to $-0.5$ eV, and the completely unfilled Cr $d$ orbitals lie in the range of 0.7 eV to 4.2 eV, leading to an $E_g$ of 1.2 eV.
This indicates a high spin Cr$^{3+}$ configuration of $S=\frac{3}{2}$, leading to the total moment of $3\mu_B$.
The moment is mainly contributed by Cr (2.9 $\mu_B$), with a smaller contribution from Se2 (0.1 $\mu_B$) and negligible contributions from the remaining ions.
Note that the exchange splitting of the $t_{2g}$ manifold is about 2.5 eV.

\begin{figure}[tb] 
\centering
\includegraphics[width=\columnwidth]{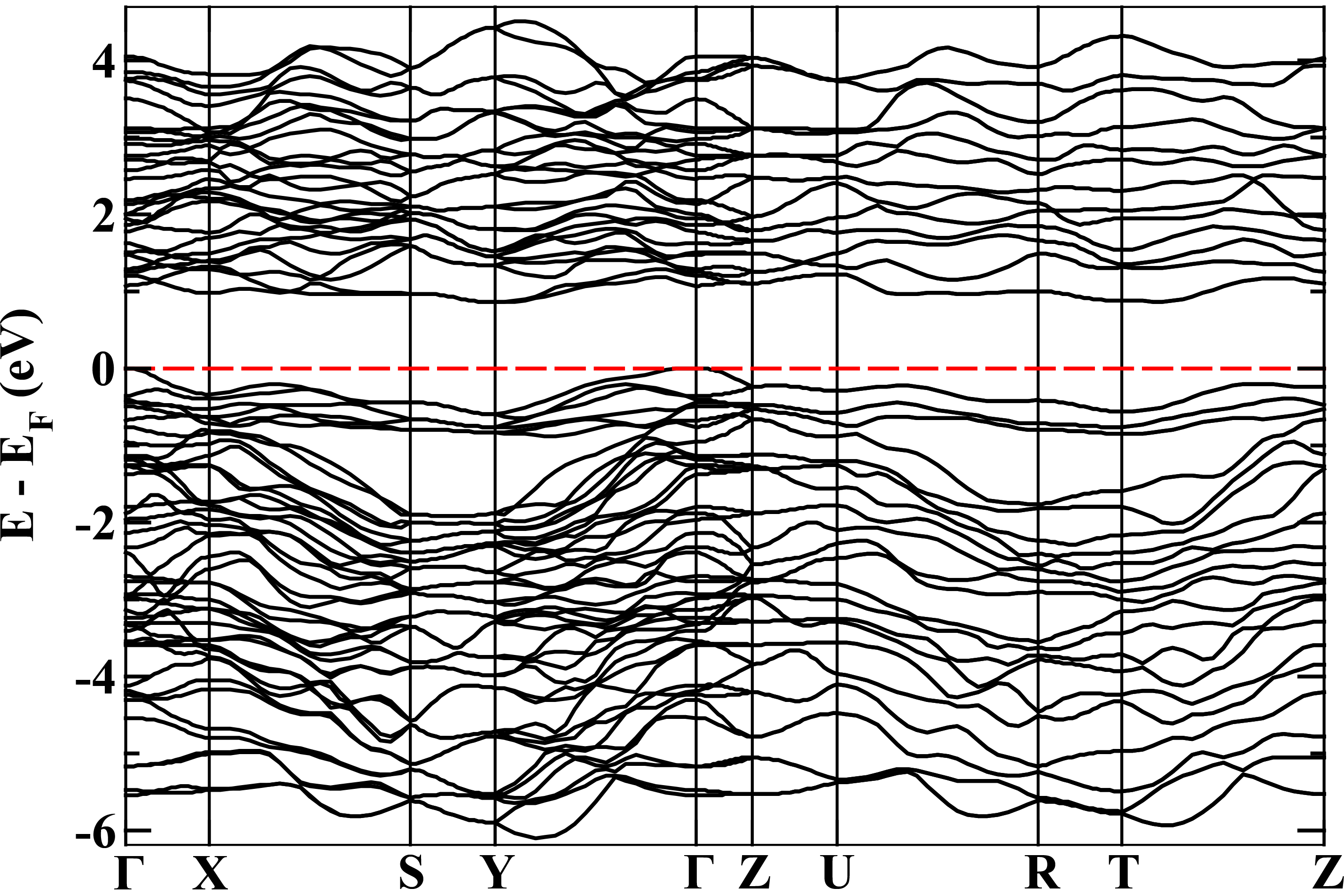}
\caption{AFM band structure of \css~at $d_{\rm Cr-Cr}=3.39$ \AA.
The band sticking is also observed at the top of BZ along the $Z-U-R-T-Z$ line, as in Fig. \ref{fm_band}.
}
\label{afm_band}
\end{figure}

\begin{figure}[t] 
\centering
\includegraphics[width=\columnwidth]{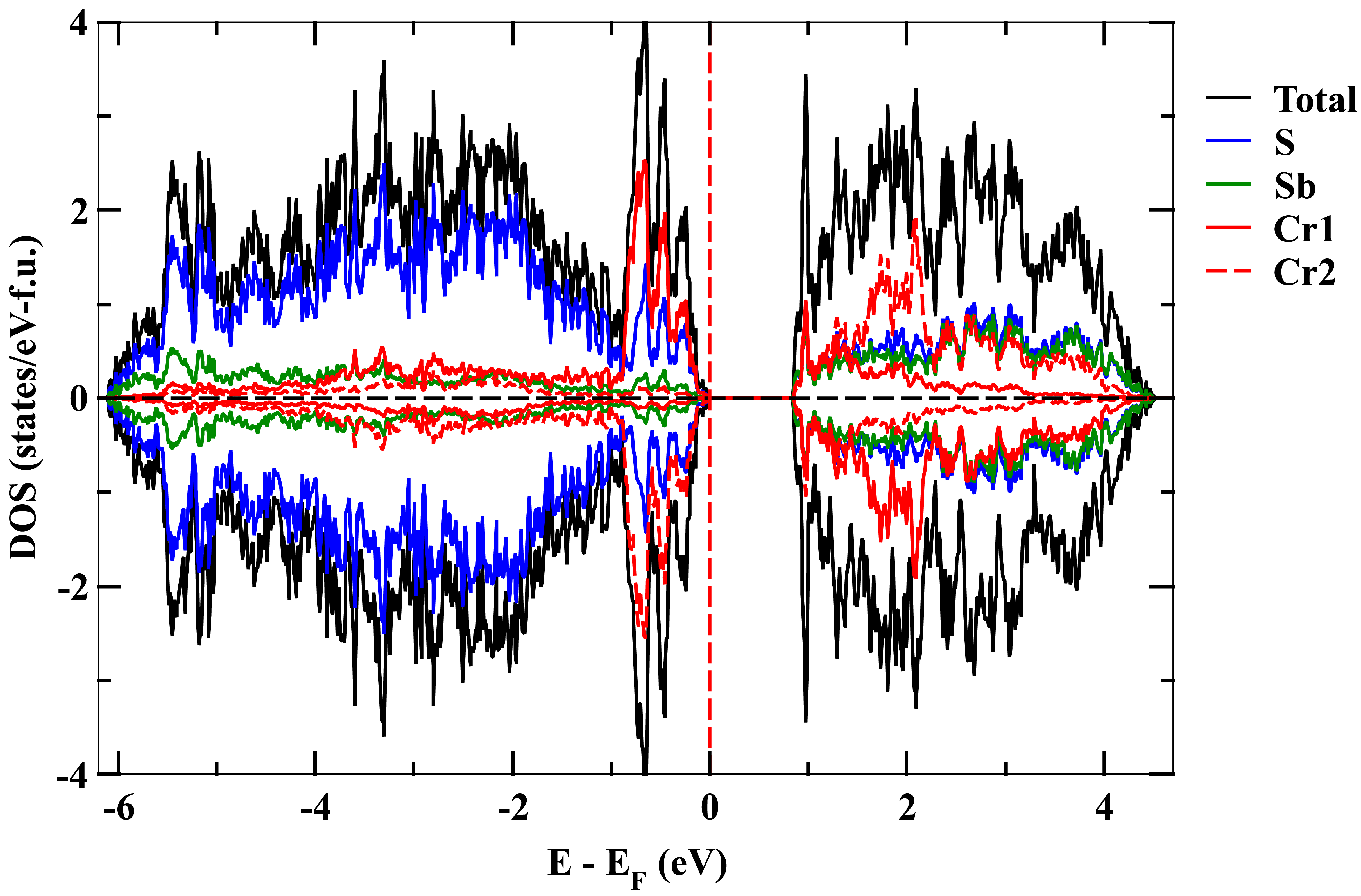}
\caption{AFM total and atom-projected DOSs of \css~
at $d_{\rm Cr-Cr}=3.39$ \AA.
The fully filled $t_{2g}$ manifold above $-0.9$ eV shows three peaks clearly, characteristic of the quasi-1D system.
}
\label{afm_dos}
\end{figure}

{\it AFM state of \rm \css.}
For the compound, the FM is energetically favored slightly over the AFM by about 8 meV/f.u. at the optimized $d_{\rm Cr-Cr}=3.48$ \AA.
In contrast, at the presumed experimental $d_{\rm Cr-Cr}=3.39$ \AA,  
the AFM is favored by about 12 meV/f.u.
So the magnetic ground state is very sensitive to $d_{\rm Cr-Cr}$ of this range, as shown in the top panel of Fig. \ref{energy},
indicating that \css~lies in the vicinity of the critical region of the magnetic phase transition.
This may account for controversial experimental observations on the magnetic ground state of \css~\cite{odink,volkov,c.li2025},
requiring detailed experimental investigations of the crystal structures.
The energy difference yields the superexchange $J'=2.5$ meV$\approx30$ K, sizable to the experimental $T_N$ \cite{c.li2025}.

We discuss the electronic structure of the AFM state observed recently,
since the FM one is very similar to that of \csse~(not shown here).
The band structure and corresponding DOSs of AFM \css~are given in Figs. \ref{afm_band} and \ref{afm_dos}, respectively, at $d_{\rm Cr-Cr}=3.39$ \AA.
The fully filled $t_{2g}$ majority manifold resides right below $E_F$
with a narrow bandwidth of 0.9 eV, while the bottom of the conduction band is dominated by the $e_g$ manifold.
This leads to the $d-d$ $E_g$ of 0.8 eV, consistent with the experimental value \cite{c.li2025}.
The dispersion of the narrow $t_{2g}$ majority along the $\Gamma-Y$ line
provides $t_b\approx0.2$ eV,
roughly $\frac{3}{2}$ of the value in the FM \csse.
In this state, the Cr local moment is 2.7 $\mu_B$ and the moments of the other ions are tiny, 0.03-0.07$\mu_B$.

\begin{figure}[tb] 
\centering
\includegraphics[width=\columnwidth]{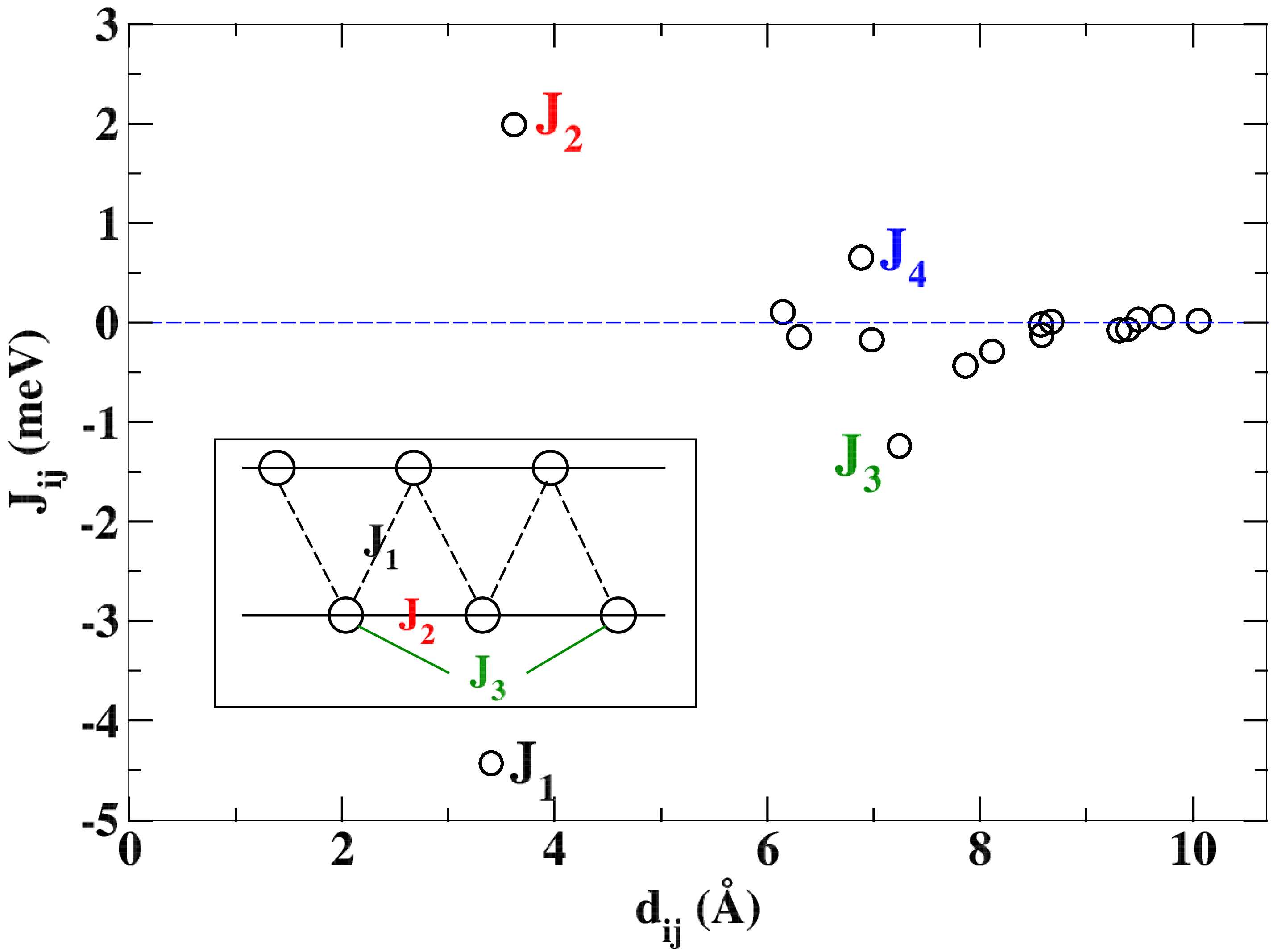}
\caption{Exchange parameters at $d_{\rm Cr-Cr}=$3.39 \AA~ in \css.
The intra-chain interactions of $J_1$, $J_2$, and $J_3$ are depicted in the Inset,
and $J_4$ indicates the NN inter-chain interaction.
Here, $d_{ij}$ denotes the inter-ionic distance between two Cr ions.
Inset: Double-rutile chains along the $\hat{b}$ axis,
resulting in the zigzag Cr-spin chains for the AFM \css.
}
\label{j1}
\end{figure}

\subsection{Magnetic interactions}

To unveil the origin of the Bethe-Slater curve-like magnetic transition, 
we investigate the magnetic exchange parameters $J_{ij}$ in the range $d_{\rm Cr-Cr}=$3.39 \AA~to 3.66 \AA, covering all experimentally and theoretically observed values for both compounds.
The parameters $J_{ij}$ are calculated for the classical spin Hamiltonian $\mathcal{H}$ = $-\sum_{i,j}J_{ij}\mathbf{e}_i\cdot\mathbf{e}_j$ as implemented in {\sc tb2j} code \cite{tb2j}, 
where $\mathbf{e}_i$ = $\frac{\mathbf{S}_i}{|S_i|}$ is the normalized spin vector.
Here, a negative $J_{ij}$ value corresponds to an AFM interaction between the $i$-th and $j$-th Cr ions.
(For simplicity, we denote $J_{ij}$ by $J_{j}$ below for a fixed $i$-th Cr ion.)

Figure \ref{j1} shows the parameters versus the ionic distance $d_{ij}$ for \css~at $d_{\rm Cr-Cr}=$3.39 \AA. 
(The results for $d_{\rm Cr-Cr}=$3.48 for \css, and 3.60 and 3.66 for \csse, in units of \AA, are given in SM.)
The dominant parameters are the three intra-chain interactions $J_1$, $J_2$, and $J_3$,
and the NN inter-chain interaction $J_4$, 
while all other parameters are negligible.
The NN intra-chain interaction $J_1$ is mediated by Se or S ions on the $ac$ plane, 
whereas the second- and third-NN intra-chain interactions, $J_2$ and $J_3$,
arise from direct interactions between Cr spins along the $\hat{b}$ axis,
as shown in the Inset of Fig. \ref{j1}.
The distance between Cr spins associated with $J_1$ is slightly shorter (by around 0.1 \AA) than that corresponding to $J_2$, 
while those associated with $J_3$ and $J_4$ are also comparable to each other and are about twice as large as the distances for $J_1$ and $J_2$.
Specifically, the distances corresponding to $J_2$ and $J_3$ are equal to the $b$ lattice parameter and twice this parameter, respectively.

\begin{figure}[t] 
\centering
\includegraphics[width=\columnwidth]{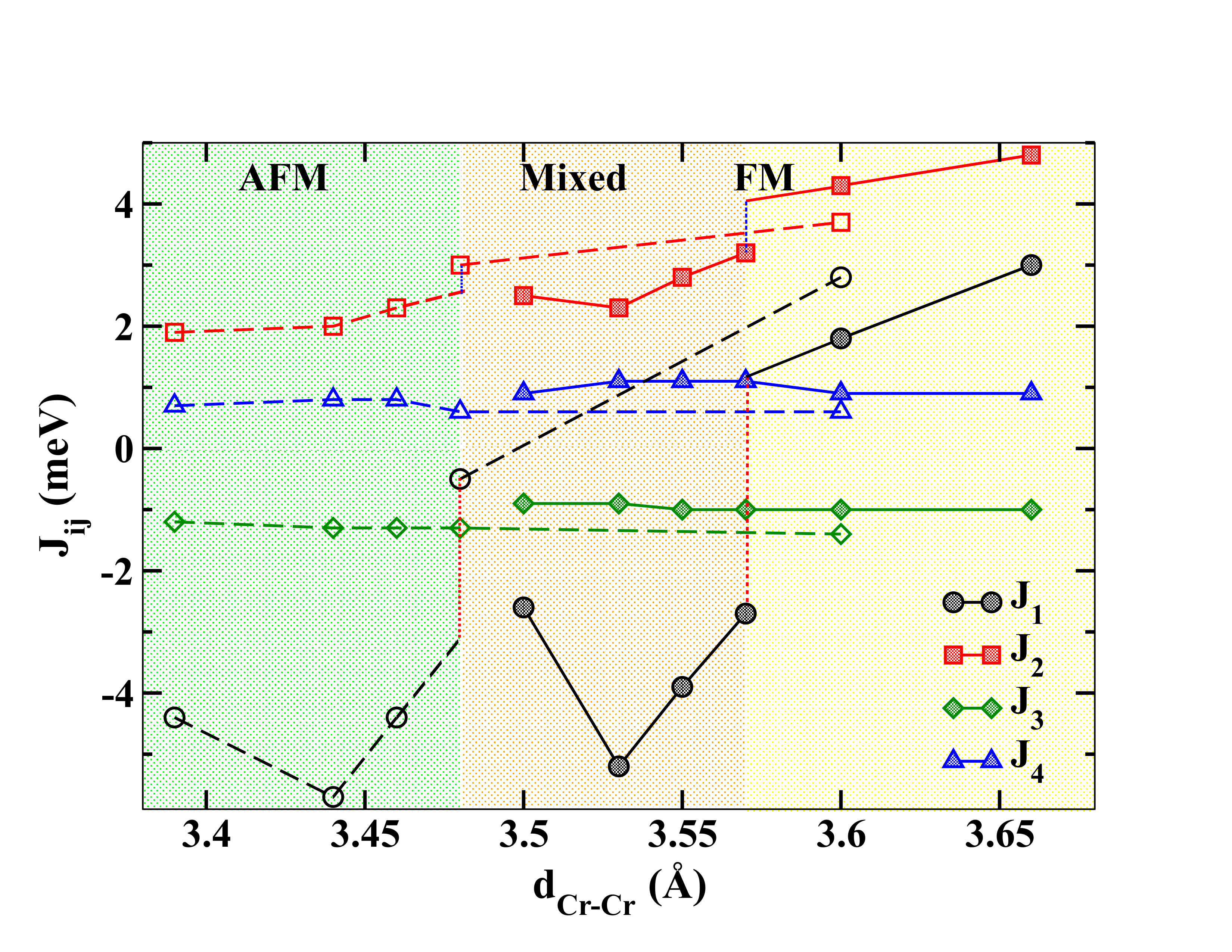}
\caption{Variations of dominant exchange parameters, in the range of $d_{\rm Cr-Cr}$=3.39 \AA~to 3.66 \AA.
The data connected by the solid and dashed lines represent \css~and \csse, respectively.
In the mixed region of $d_{\rm Cr-Cr}$=3.48 \AA~to 3.58 \AA,
the FM state is preferred to the AFM state for \css,
whereas the AFM state is more favored over the FM state for \csse.
Remarkably, at the critical value of each system, $J_1$ and $J_2$ show discontinuities, which are much more pronounced in $J_1$, as denoted by dotted-lines.
}
\label{j_sum}
\end{figure}

As displayed in Fig. \ref{j1}, the largest interaction is AFM $J_1=-4.8$ meV.
Rather unexpectedly, the second Cr–Cr direct interaction $J_3$ is AFM, 
yet this is at less than half of the NN FM direct interaction $J_2$ and is therefore largely suppressed, resulting in FM alignment along the $\hat{b}$-axis.
This behavior along the $\hat{b}$-axis remains consistent over all ranges $d_{\rm Cr-Cr}$ studied here, as given in Fig. \ref{j_sum}. 
The small positive $J_4$ indicates a relatively weak inter-chain interaction,
as expected from the quasi-1D magnets.
This indicates that the experimentally observed C-type AFM structure in \css~\cite{c.li2025} is attributed to the competition between the AFM $J_1$ and the FM $J_2$.

Figure \ref{j_sum} illustrates how the four dominant exchange parameters evolve with varying $d_{\rm Cr-Cr}$ in the range of 3.39 \AA~to 3.66 \AA.
$J_3\approx-1$ meV and $J_4\approx0.8$ meV remain nearly unchanged over the range for both compounds.
By contrast, as $d_{\rm Cr-Cr}$ increases, the $J_2$ values increase smoothly, with small jumps near the boundaries at $d_{\rm Cr-Cr}\approx3.48$ \AA~for \css~and 3.58 \AA~for \csse, and remain ferromagnetic throughout,
although the distance between Cr ions corresponding to $J_2$ remains fixed.
Specifically, $J_2\approx2$ meV for \css~at $d_{\rm Cr-Cr}=3.39$ \AA~and $J_2\approx4$ meV for \csse~at $d_{\rm Cr-Cr}=3.60$ \AA,
consistent with a relation of $J_2\propto t^2_b$.
Here, the hopping parameters $t_b$ are obtained for FM \csse~and AFM \css~at the corresponding $d_{\rm Cr-Cr}$ values, which represent their respective experimental structures, in Sec. III \ref{el_str}. 
The most significant variation is observed in $J_1$,
since such a slight change in $d_{\rm Cr-Cr}$ scarcely 
affects the Cr–Cr distances associated with the remaining $J_i$ interactions and local environments around the Cr ions.
As $d_{\rm Cr-Cr}$ increases, 
the $J_1$ values display an evident discontinuity 
where the transition from AFM to FM occurs at $d_{\rm Cr-Cr}\approx3.48$ \AA~for \css~and $d_{\rm Cr-Cr}\approx3.58$ \AA~for \csse.
In the AFM-favored region for each compound,
the $J_1$ values vary approximately parabolically, 
reaching minima of  $-5.8$ meV for \css~and $-5.3$ meV \csse.
Then, the $J_1$ values abruptly jump 
at the transition boundaries and subsequently increase linearly in the FM-favored region.

Considering contributions of $J'_iS^2$ for each spin pair and neglecting higher order $J'_i$ interactions,
the energies of these systems are given by
$E_{\rm FM}\approx-(J'_1+J'_2)S^2$ and $E_{\rm AFM}\approx(J'_1-J'_2)S^2$. 
This leads to an energy difference of $\Delta E\approx2J'_1S^2=2J_1$, where $J'_i=J_i/S^2$.
The energy difference shown in the top panel of Fig. \ref{energy} follows this relation reasonably well 
in the AFM-favored cases near the boundary of the magnetic phase transition.
Consequently, the AFM $J_1$ interaction plays a crucial role in stabilizing the AFM states.
On the other hand, a simple Ising model with only NN exchange interactions in a 1D $S=\frac{3}{2}$ system leads to the energy difference between FM and AFM states of $\Delta E=2J_2$.
For the FM-favored cases near the transition boundary,
the resulting estimate of $J_2$ is about half of the value shown in Fig. \ref{j_sum}, indicating that both FM $J_1$ and $J_2$ interactions are essential for stabilizing the FM states.

\section{Discussion}
As mentioned in the Introduction, superconductivity emerges at $P\sim33$ GPa in \csse~that is an insulating FM at ambient pressure.
Using the experimental structure including the internal parameters at $P=22.4$ GPa \cite{c.li2024}, 
we calculated the magnetic ground state. 
At the pressure ($d_{\rm Cr-Cr}=3.18$ \AA), AFM has a lower energy than in FM by 12.7 meV/f.u., following the Bethe-Slater-curve-like behavior observed in Sec. III \ref{energetics}.
At the AFM state, the Cr local moment of 0.9 $\mu_B$, whereas those of the other ions are minor,
indicating a pressure-induced spin-state transition from nominally $3d^{3\uparrow}$ to $3d^{2\uparrow,1\downarrow}$.
This state is itinerant, resulting from a substantial volume collapse of approximately 22\% relative to ambient pressure.
This indicates a transition of an insulating FM$\rightarrow$itinerant AFM$\rightarrow$superconductivity with increasing pressure, 
suggesting that the superconductivity is likely associated with AFM fluctuations rather than FM ones.
At the superconducting phase, the Fermi surfaces show very strong 1D character (see SM), implying pronounced nesting instabilities along the chain direction, reminiscent of the high $T_c$ cuprate Ba$_2$CuO$_{3+\delta}$ \cite{hsjin}.
These issues require further study. 

Another notable feature is the presence of a second sharp peak at $T_2=94$ K in the magnetic susceptibility of \css, accompanied by a corresponding peak in the heat capacity data, in addition to the peak observed at $T_N=43$ K \cite{c.li2025}, as mentioned in the Introduction.
Li {\it et al.} fitted the susceptibility data and obtained the effective moments of 3.74 and 2.76 (in units of $\mu_B$/Cr) 
at the region of $T \ge T_2$ and $T_N\le T\le T_2$ \cite{c.li2025}, respectively. Each value is close to the theoretical value for
Cr$^{3+}$ ($3d^3$, $S=\frac{3}{2}$) and Cr$^{4+}$ ($3d^2$, $S=1$). Additionally, the $g$-factor derived from the ESR spectra is 1.96 at $T_2$, shows a jump of 0.02 at $T_2$, and then increases smoothly to 1.99 at $T_N$ upon cooling.
The latter $g$-factor corresponds to Cr$^{3+}$ ($t_{2g}^3$), for which only a tiny spin–orbit coupling is expected.
Therefore, they concluded that this system undergoes a charge-transfer transition from Cr$^{3+}$ ($3d^3$) to Cr$^{4+}$ ($3d^2$) at $T_{CT}(=T_2)$, rather than a spin-state transition, returning to Cr$^{3+}$ at $T_N$.
Although no change in crystallographic symmetry is observed at $T_{CT}$ and $T_N$, the $T$-dependent variations show discontinuities at $T_{CT}$ in the lattice parameters of $a$ and $c$, whereas the $b$ parameter remains unchanged.
Remarkably, the $a$ and $c$ show opposite thermal expansion behaviors, positive and negative, respectively, in the range of between $T_{CT}$ and $T_N$, accompanying small discontinuities in the average Cr-S bond length at both temperatures.   
In the $T$ range, these variations lead to a little volume collapse and a more sudden drop in the ratio of $\frac{c}{a}$, 
but less than 1\% in both. 
Below $T_N$, the variations for all lattice parameters are negligible.

To investigate the charge transfer transition,
the atomic internal parameters were optimized while keeping the experimentally observed lattice parameters at around $T_{CT}$, 
with a unit cell volume less than 1\% smaller than that at room $T$ \cite{c.li2025}.
However, in the optimized structure the local structural parameters, such as atomic distances and bond angles, remain identical (with a precision of 0.01 \AA) to those obtained from the room-$T$ lattice parameters.
So, as one may expect, no evident indications for the charge-transfer transition are found within our computational accuracy, even when explicit correlation effects are included via the DFT+U approach that has often been shown to capture such behavior \cite{wep2012}.
Thus, to elucidate the unusual $T$-dependent behaviors 
in the intermediate temperature region $T_N\le T\le T_{CT}$,
approaches beyond the conventional DFT would be necessary.
In addition, for {\it ab initio} theoretical investigations, complementary experimental measurements capable of probing the $T$-dependence of local structural parameters seem to be essential.
More fundamentally, direct experimental measurements are necessary to confirm the previously proposed charge-transfer transition.
Therefore, these issues warrant further investigation from both theoretical and experimental viewpoints.

\section{Summary}
Throughout the {\it ab initio} calculations performed with the accurate all-electron full-potential code {\sc wien2k}, we investigated the origin of the distinct magnetic ground states for the quasi-1D ternary chromium trichalcogenides CrSb$X_3$ ($X$=S, Se), composed of infinite double-rutile chains of edge-sharing Cr$X_6$ octahedra.
In contrast to previous reports showing substantially underestimated $E_g$ \cite{y.xun2021,g.wang2021}, 
our calculations within GGA well reproduce insulating states with $E_g$'s very close to the experimental values, indicating that these are band insulators rather than Mott insulators.

Remarkably, our results show that 
changes in the Cr-Cr bond length $d_{\rm Cr-Cr}$ lead to variations in the intra-chain magnetic interactions: the  shortest Cr-Cr direct exchange interaction $J_2$ along the chain direction $\hat{b}$ and even more pronounced variations in the intra-chain NN superexchange interaction $J_1$ along an oblique in-plane direction.
At the critical value $d^c_{\rm Cr-Cr}\approx3.53(\pm0.05)$ \AA,
evident discontinuities emerge in both the energy difference between AFM and FM and in $J_1$, 
indicating the first-order magnetic phase transition. 
These findings suggest that \css~lies in close proximity to the critical region of the phase transition, in apparent agreement with previously reported, yet controversial, experimental observations of its magnetic state \cite{odink,volkov,c.li2025}. In contrast, \csse~remains robustly ferromagnetic, consistent with the experiments \cite{odink,volkov}.

Taken together, the observed bond-length dependence--extending to a broader class of quasi-1D Cr-based magnets--reveals an emergent Bethe–Slater-like behavior driven by competing exchange pathways in a quasi-1D transition-metal system,
in particular with the critical value $d^c_{\rm Cr-Cr}$.
This establishes bond length as a decisive control parameter for magnetic order
in these systems, as often observed in transition metal-based compounds.
However, for these systems it would be much easier to access and tune multiple magnetic phases by bond-length engineering,
since these have (quasi)-one-dimensionality and lie in close proximity to the critical region.

\section*{Acknowledgments}
We acknowledge useful communications 
with Kyo-Hoon Ahn, Hyun-Yong Lee, and Kyujoon Lee for magnetic transitions.
This research was supported by National Research Foundation of Korea (NRF) Grant (RS-2024-00392493).

\vskip 8mm
{\bf End of Paper}

\clearpage
\newpage\onecolumngrid
\renewcommand\thefigure{S\arabic{figure}} 
\newcounter{myfig} 
\setcounter{figure}{0}

\renewcommand\thetable{S\arabic{table}} 
\newcounter{mytable} 
\setcounter{table}{0}

\section{Supplemental Material}


This Supplemental Material provides additional information supporting the main text. The information is described in the figure captions.
\vskip 8mm

\begin{figure*}[!ht] 
\includegraphics[scale=0.35]{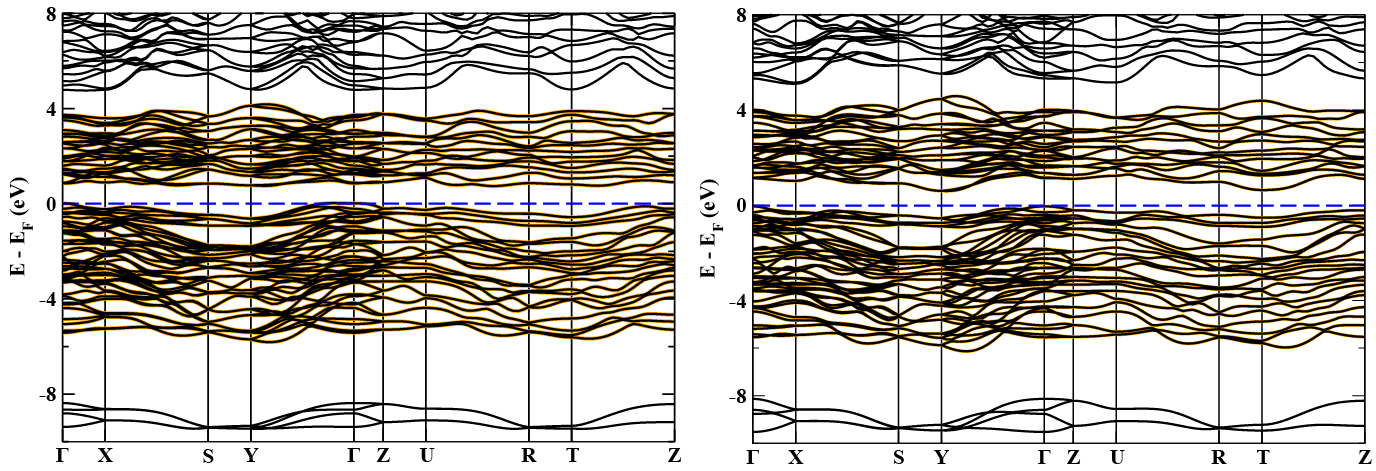}
\caption{Comparisons of the Wannierized bands (brown symbols) with the {\it ab initio} AFM bands for (left) \csse~ and (right) \css, showing the very good agreement over a 10 eV span.
}
\label{wf}
\end{figure*}

\begin{figure*}[!ht] 
\includegraphics[scale=0.30]{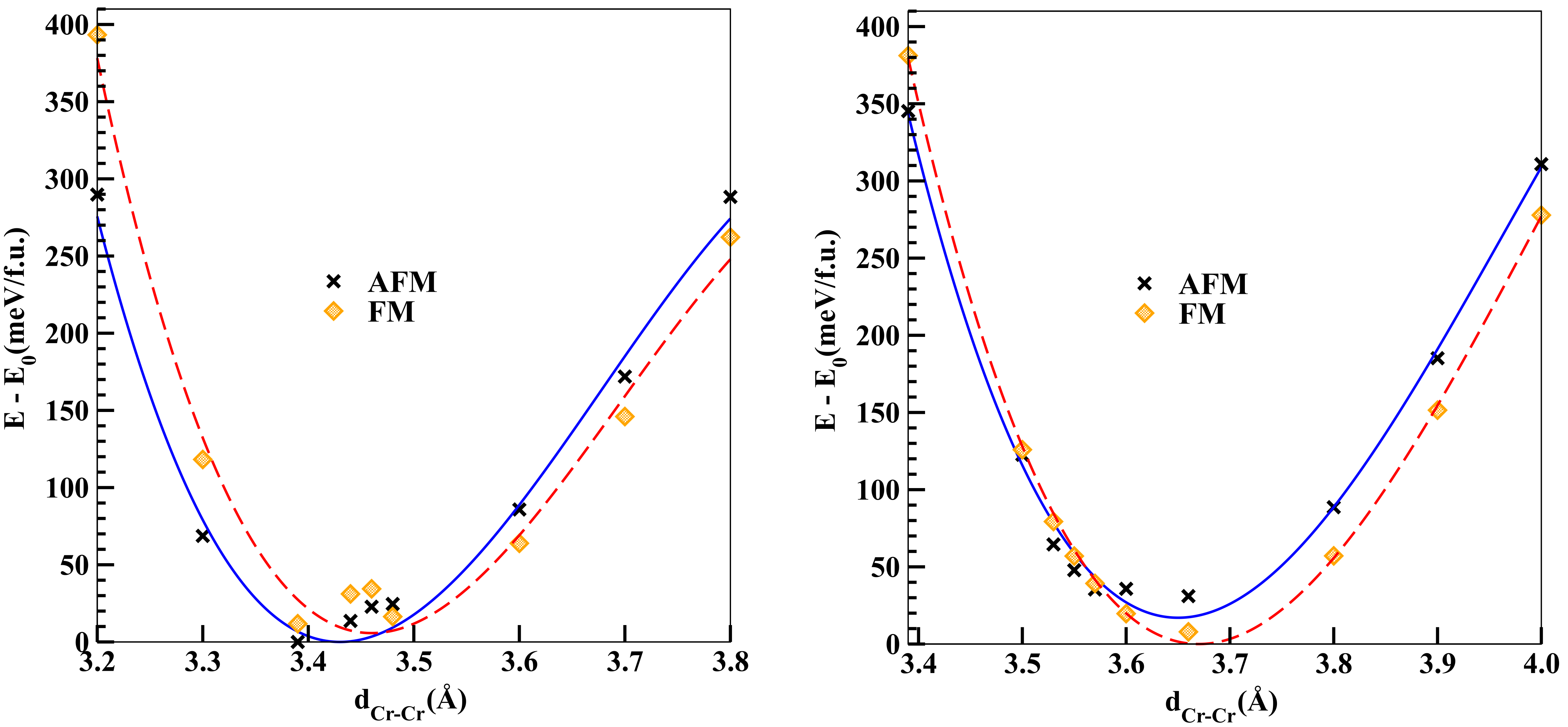}
\caption{Variations of energies in AFM and FM, as varying the Cr-Cr bond lengths $d$, for (left) \css~ and (right) \csse.
These variations can be well fitted by a function that includes terms up to cubic order, $\varepsilon_0+a_1(d-d_0)^2+a_2(d-d_0)^3$, as described by the solid (AFM) and dashed (FM) lines.
For \css, $d_0=3.43 (3.46)$ \AA, $\varepsilon_0= 0 (5.7)$ meV, $a_1= 3981.7 (4029.9)$ meV/\AA$^2$, $a_2= -5346.2 (-5690.7)$ meV/\AA$^3$ in the AFM (FM) state.
For \csse, $d_0=3.65 (3.67)$ \AA, $\varepsilon_0=17.0 (0)$ meV, $a_1= 3787.0 (3788.7)$ meV/\AA$^2$, $a_2= -4011.8 (-3766.4)$ meV/\AA$^3$ in the AFM (FM) state.
As usually observed, the equilibrium distance $d_0$ is nearly independent of the choice of magnetic state.
Here, $E_0$ denotes the minimum energy for each compound.
}
\label{varying_J}
\end{figure*}

\begin{figure*}[!ht] 
\includegraphics[scale=0.26]{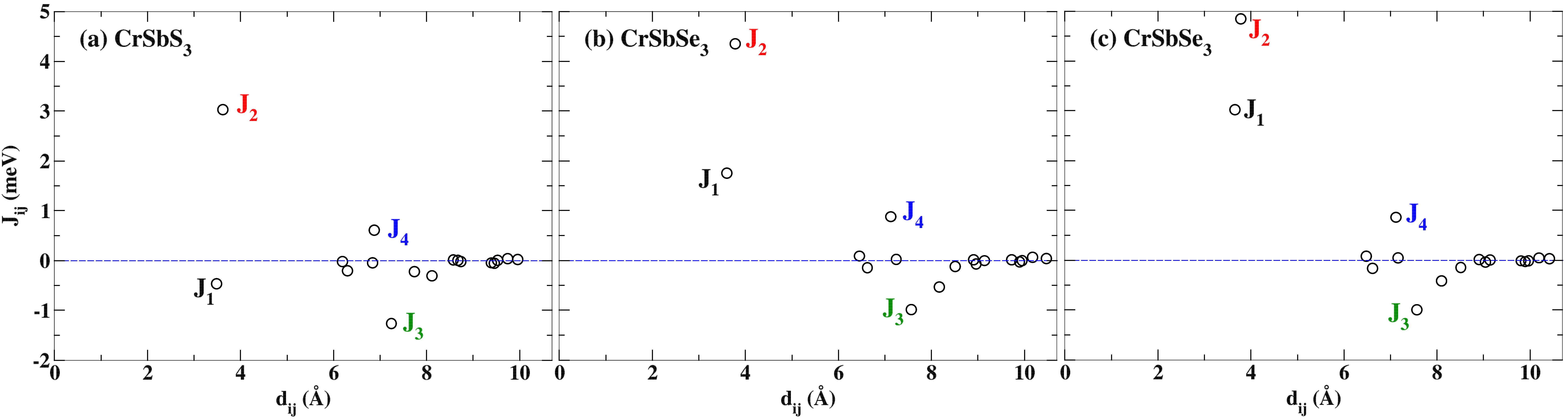}
\caption{Exchange parameters (a) at $d_{\rm Cr-Cr}=$3.48 \AA~ in \css, (b) and (c) at $d_{\rm Cr-Cr}=$3.60 \AA~ and 3.66 \AA~ in \csse.
(For the symbols of $J_i$'s, see the main text.)
}
\label{varying_J}
\end{figure*}

\begin{figure*}[!ht] 
\includegraphics[scale=0.18]{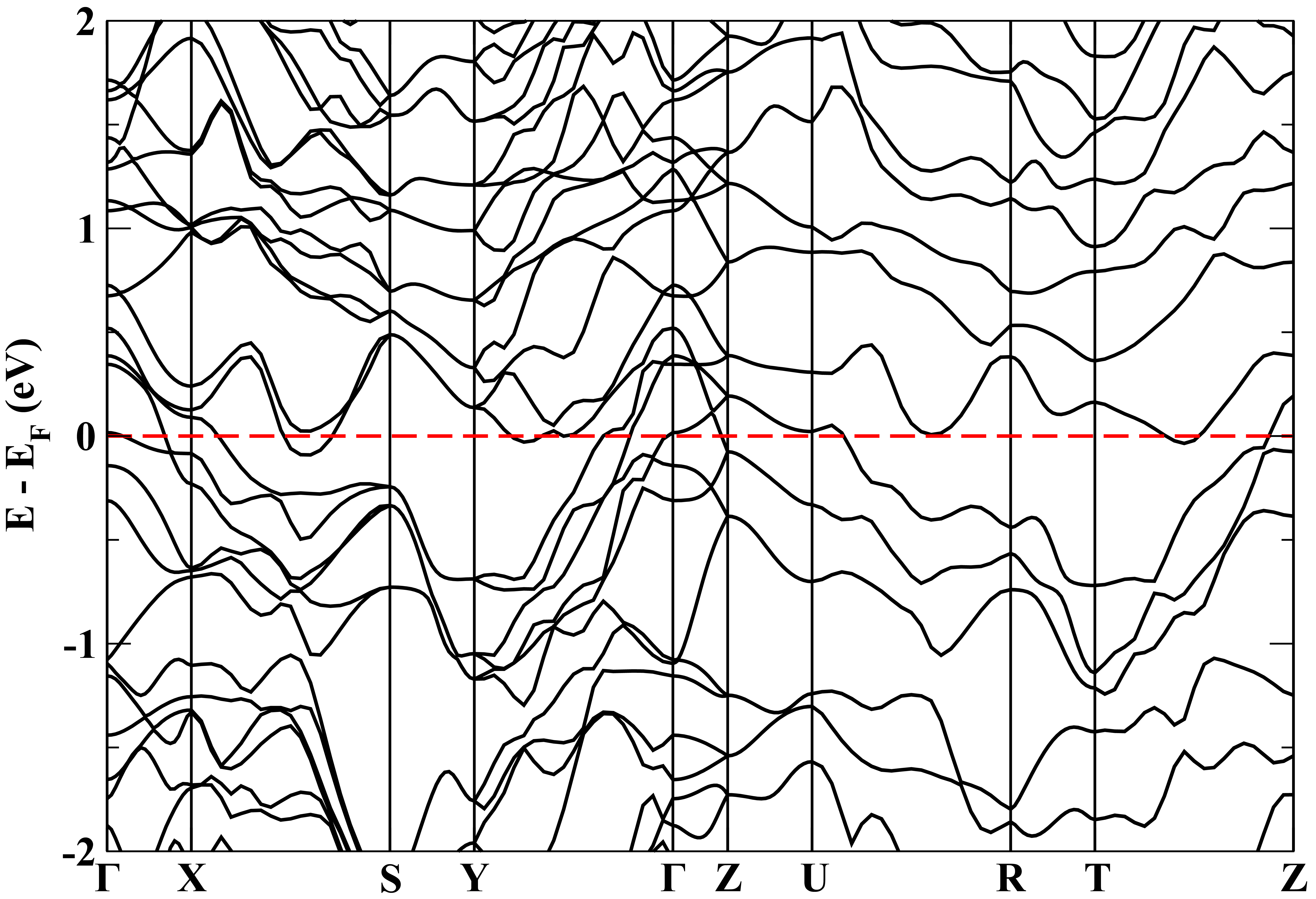}
\caption{Enlarged band structure 
in the AFM \csse~ at 22.4 GPa, in the range of $-2$ eV and 2 eV.
In the AFM state, the Cr local moment is about 0.9 $\mu_B$.
}
\label{csse_p_full}
\end{figure*}

\begin{figure*}[!ht] 
\includegraphics[scale=0.35]{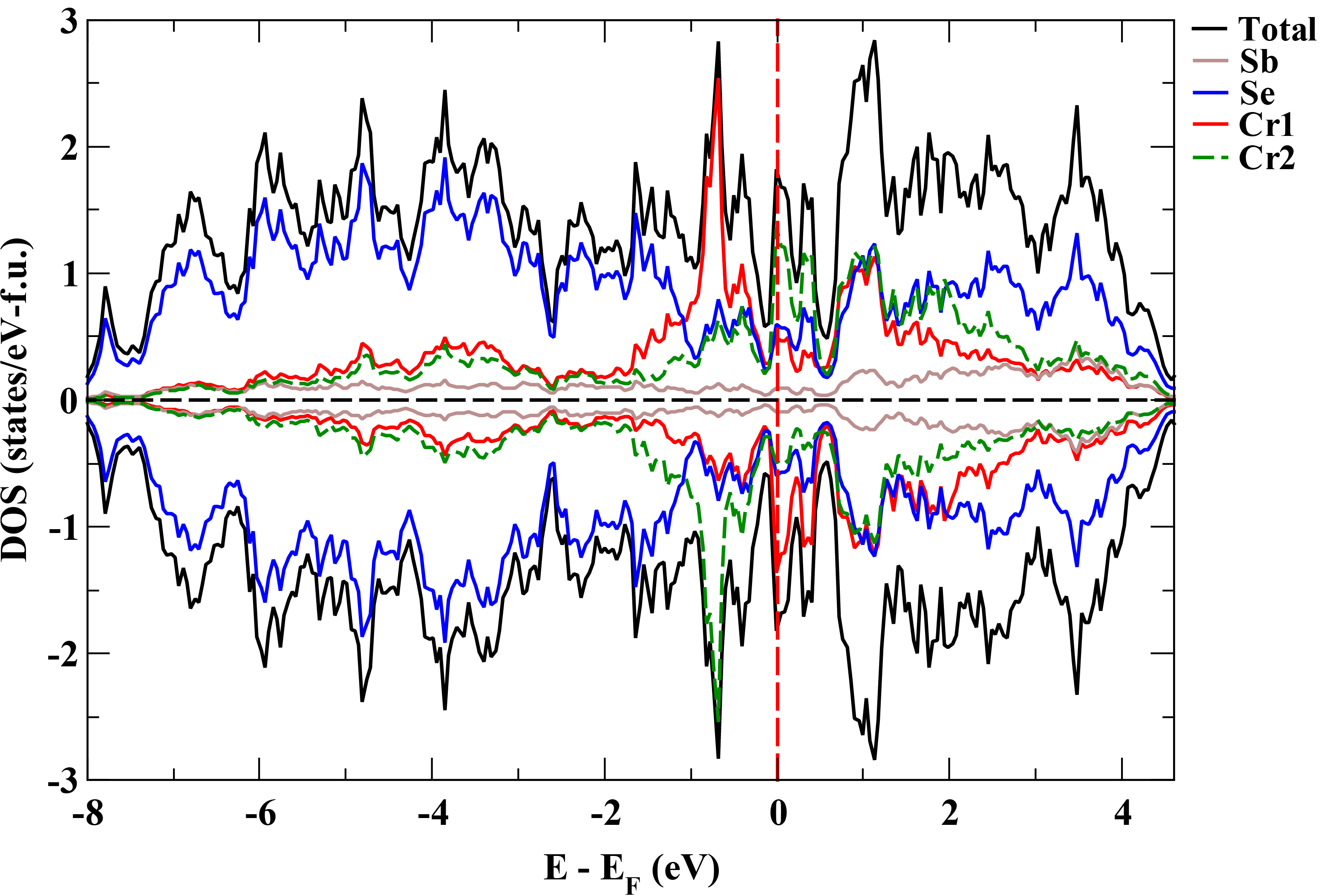}
\caption{Total and atom-projected densities of states in the AFM \csse~ at 22.4 GPa, in the full energy range,
indicating the Cr$^{3+}$ ions with a low spin $S=1$ configuration  $t_{2g}^{2\uparrow,1\downarrow}$ (see in the region of $-1.5$ eV to 0.5 eV.) 
}
\label{csse_p_dos}
\end{figure*}

\begin{figure*}[!ht] 
\includegraphics[scale=0.35]{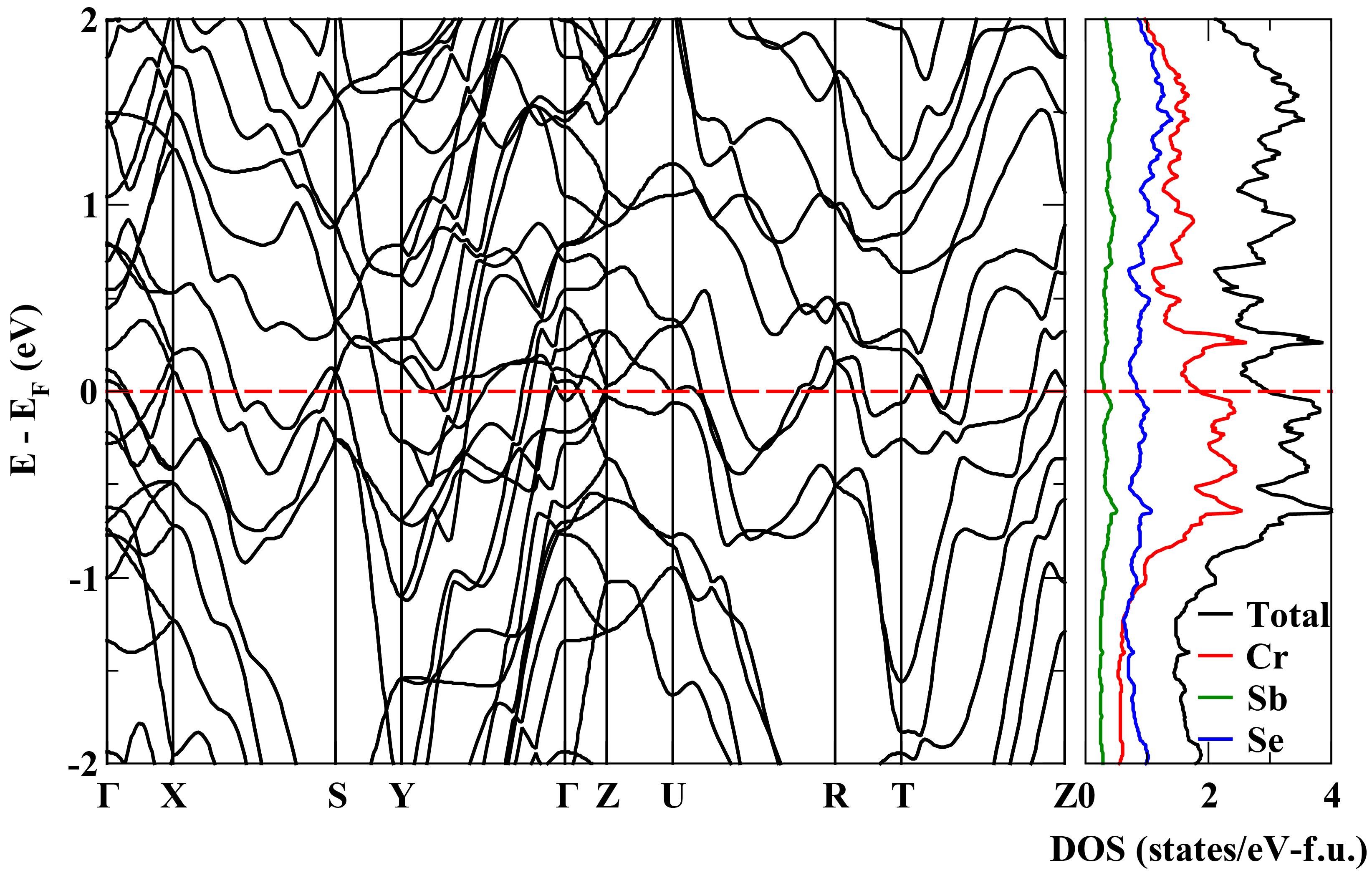}
\hskip 10mm
\includegraphics[scale=0.45]{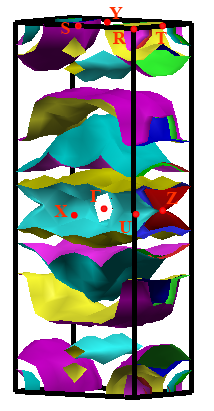}
\caption{Electronic structure and Fermi surfaces in the nonmagnetic \csse~ at 40 GPa, where the superconductivity emerges.
Left: enlarged band structure and total and atom-projected DOSs. 
The DOS at the Fermi energy $E_F$ is about 2.9 states/eV per f.u., 
with approximately 65\% attributable to Cr ions.
Right: Fermi surfaces that display strong 1D character, suggesting a  considerable 1D instability along the chain direction ($\hat{b}$-axis). 
Here, the experimentally observed lattice parameters and internal parameters were used \cite{c.li2024}.
}
\label{csse_sc}
\end{figure*}

\end{document}